%Paper: hep-th/9404124
%From: jwb@maths.nottingham.ac.uk (Dr John W. Barrett)
%Date: Wed, 20 Apr 94 16:41:37 +0100
%Date (revised): Fri, 22 Apr 94 15:48:55 +0100
%Date (revised): Fri, 22 Apr 94 16:30:30 +0100
%Date (revised): Wed, 21 Sep 94 18:16:31 +0100

%hep-th/9404124
%
%                   First order Regge calculus
%
%                       John W. Barrett
%
%
% amstex v2.1
%
%
\documentstyle{amsppt}
%\refstyle{B}
\NoBlackBoxes
\magnification=\magstep 1
\def\today{21 September 1994}

\topmatter
\title First order Regge Calculus\endtitle
\author John W. Barrett\endauthor
\date\today\enddate
\address
Department of Mathematics,
University of Nottingham,
University Park,
Nottingham,
NG7 2RD, UK
\endaddress
\email John.Barrett\@nottingham.ac.uk \endemail

\abstract
A first order form of Regge calculus is defined in the spirit of Palatini's
action for general relativity. The extra independent variables are
the interior dihedral angles of a simplex, with conjugate variables
the areas of the triangles.

There is a discussion of the extent
to which these areas can be used to parameterise the space of edge lengths
of a simplex.
\endabstract

\endtopmatter

\document

\def\d{\hbox{d}}
\def \squarebox{$\sqcap\kern-1.5ex\sqcup$} %end-of-proof symbol

\def\args{\left(\sigma^4,\sigma^2\right)} %commonly used argument
\def\argsij{\left(\sigma^4,\sigma^2_{ij}\right)} %commonly used argument
\def\pderiv#1#2{{\partial#1\over\partial#2}} %partial 1 by partial 2
\def\deriv#1#2{{\d#1\over\d#2}} %derivative of 1 by 2
\def\modsigma#1 {|\sigma#1|}   %volume of simplex
\def\R{{\Bbb R}}  % the real numbers
\def\metric{\operatorname{metric}}

Regge's equations of motion for Regge calculus \cite{Regge 1961},
a discrete version of general relativity, are derived
from an intuitively appealing idea about the correct form for an
action principle. However the equations are a little
complicated, each equation involving a fairly large number of
neighbouring edge lengths in a complicated pattern, and involving
combinations of polynomials, square roots, and arccosines of these
edge lengths. It is desirable from many points of view to understand
these equations more fully, and maybe simplify their form.
For example, the second order nature
of the equations makes implementation of the Cauchy problem for
numerical relativity rather complicated \cite{Sorkin 1975, Tuckey 1993}.

In quantum gravity, models for three space-time dimensions have been
constructed, either involving a path integral \cite{Witten 1988} or in a
discrete
version as inspired by Regge and Ponzano \cite{1968}.
Witten's construction starts with a
first order action for gravity. Regge and Ponzano's model has semiclassical
limits which involve generalisations of Regge calculus to degenerate metrics
\cite{Barrett and Foxon 1994}.
The appearance of degenerate metrics is the way in which first order actions
for general relativity differ, in a physical sense, from the usual second order
Einstein-Hilbert action.  These considerations of models of quantum gravity
are the main motivation for studying first order actions for Regge calculus.

The Regge action involves computing defect angles, which are
$2\pi$ minus the sum of dihedral angles. A dihedral angle is the
angle between two different faces (3-simplexes) in a 4-simplex. This paper
considers the possibility of extending the Regge action by
taking the dihedral angles to be independent variables, rather
than as just functions of the edge lengths. As far as simplifying or
radically restructuring the Regge calculus is concerned, this work has
to be regarded as preliminary. It hints at a calculus in which angles
play a dominant role.

The simplest signature for the metric to consider is the case of the positive
definite metric, in which case the dihedral angles are angles in the ordinary
sense of Euclidean geometry. However, it is perfectly possible to consider
other signatures. This paper will discuss mainly the Euclidean case.
The phenomena associated with Lorentzian angles in Regge calculus are
discussed in \cite{Sorkin 1975} and \cite{Barrett and Foxon 1994}.
Also, the dimension of the manifold is taken to be four throughout, for
familiarity, but similar constructions may be made in other dimensions.

The first section is a discussion of first order actions for general
relativity and a property of the Regge calculus action which is analogous
to a property of the first order action for general relativity.
Then there is a definition of a first order action for Regge calculus with
the edge lengths and dihedral angles as independent variables, but for
which the variation has to be constrained. Using a Lagrange multiplier
for each simplex, a second action is defined in which each variable can be
varied without constraint.

Both of these new actions have as stationary points all of the stationary
points of the original Regge action, but there may be extra ones. These
extra ones may arise from the discrete ambiguity in reconstructing the
ten edge lengths for a 4-simplex from the values of the areas of the ten
triangles.

\head First Order Actions \endhead

Regge calculus is similar to the second-order formulation of general
relativity. This is its original
formulation by Einstein and Hilbert, where the metric is the only
independent variable in the action. The
idea is to introduce something similar to a first order form, where
there are more variables, but the equations are simpler, involving
only one derivative. It is similar to the idea of introducing
two first order differential equations in place of one second order equation.

In general relativity, the Einstein-Hilbert action $S$ is a
function of the metric tensor $g$. Palatini's \cite{1919} action
$P(g,\Gamma)$ is
a function of the metric and a general torsion-free, but otherwise
unrestricted connection $\Gamma$, and extends the Einstein-Hilbert
action, in the following sense
  $$
S(g) = P(g,\gamma(g))         \eqno (1)
$$
where $\Gamma=\gamma(g)$ is the unique metric-compatible connection
for $g$.

Kibble \cite{1961}, and Sciama \cite{1962}, extended the Einstein-Hilbert
action in a second way, by considering a function of the metric
and a general metric-compatible, but not torsion-free, connection.
Both of these extensions are first order actions for general
relativity, in that the action contains only first derivatives.
Both have been called Palatini formulations at times, but I prefer to
refer to the first one only as a Palatini action. The second one is
called the Einstein-Cartan-Sciama-Kibble action.

The possibility of a first order Regge calculus in
the spirit of the ECSK action was considered by Drummond \cite{1986}
and Caselle, D'Adda and Magnea \cite{1989}.
The purpose of this paper is to suggest a second avenue for constructing a
first order action, by following the Palatini formalism.

The first order actions have the property that the variational equations on
the larger spaces of fields reduce (in the absence of matter) to the
usual equations of general relativity for the metric and
connection. One has
$$
\d P={\pderiv Pg} \d g +{\pderiv P\Gamma}\d\Gamma           \eqno (2)
$$
The vanishing of the second coefficient, ${\partial
P/\partial\Gamma}$, is the variational
equation which implies that the connection is the metric
compatible one, $\Gamma=\gamma(g)$. This implies
$$ {\partial P\over\partial\Gamma}(g,\gamma(g))=0.$$

Thus, if one takes the
Einstein-Hilbert action $S$, and regards it as a function of two variables,
$g$ and $\Gamma$,
where $\Gamma$ is constrained, as in (1), then the equation of
motion in fact reduces to
$$
0=\deriv Sg = {\pderiv Pg}(g,\gamma(g)), \eqno (3)
$$
as a consequence of this.

Regge noticed a similar phenomenon to this in Regge calculus.
The action is
 $$
I(g)=\sum_{\sigma^2}|\sigma^2|\left(2\pi-\sum_{\sigma^4>\sigma^2}
\theta\args\right)                                       \eqno (4)
 $$
where the notation is as follows: $\sigma$ denotes a simplex, with the
superscript indicating the dimension, $a>b$ indicates that simplex
$a$
contains $b$ as a face, $|\sigma|$ denotes the length, area, volume etc., of
the simplex, and $\theta\args$ is the interior dihedral angle in $\sigma^4$
between
the two faces which meet at $\sigma^2$. Also, $g$ is the metric, which is
a tuple of squared edge lengths:
 $g=(\alpha_1,\alpha_2,\ldots)$, where $\alpha_k$ is the square of
the length of the k-th edge. The area of the 2-simplex and the angles
$\theta$ are
both functions of $g$.

We also need to consider the function
$J(g,\Theta)$
given by replacing each function $\theta\args$ in (4) by an
independent
variable $\Theta\args$ (equation (13) below). The symbol $\Theta$ denotes
the tuple $\left(\Theta{\args}_1,{\Theta\args}_2,\ldots\right)$ for all
pairs $\sigma^4>\sigma^2$ in the manifold. Then,
  $$
    I(g)  =  J(g,\theta(g))                         \eqno(5)
$$
Regge considered the computation
$$
\deriv Ig=\pderiv Jg +\pderiv J\Theta \pderiv\theta g       \eqno (6)
 $$
       and showed that the second term is in fact vanishing. As before, the
 notation is a little condensed: $g$ and $\theta$ are vectors, and the
second term involves multiplying a vector and a matrix together. The
situation is obviously similar to that of the \hbox{Palatini} action, and
{\it suggests} that the $\theta$'s might be considered as
independent variables.       Since
$$
\pderiv J{\Theta\args} = - \modsigma^2    \eqno (7)
$$
the second term in (6) is
$$
-\sum_{\sigma^4}\left(
\sum_{\sigma^2<\sigma^4}\modsigma^2 \pderiv{\theta\args}g
\right) \eqno (8)
$$
Regge showed \cite{1961, appendix 1} that for a fixed $\sigma^4$
$$
0=\sum_{\sigma^2<\sigma^4}\modsigma^2 \pderiv{\theta\args}g
\eqno (9)
$$
A simplified proof is presented here, as it is useful in the
       following.

 \demo{ Proof of \rm (9)} Let $i,j,$... be vertices of $\sigma^4$.
Label the faces
       of $\sigma^4$ by the vertices from $\sigma^4$ which are omitted:
thus $\sigma^3_i$ is the 3-simplex opposite vertex $i$, $\sigma^2_{ij}$ a
2-simplex, etc. Let
      $\gamma_{ij}=-\cos\theta\argsij$ for $i\ne j$,
$\gamma^{ii}=1$, and let $n_i$ be the outward unit
normal vector to face $\sigma^3_i$. Then $\sum_i\modsigma^3_i n_i=0$,
and since $\gamma_{ij}=n_i\cdot n_j$
$$
0=\sum_i \gamma_{ij} \modsigma^3_i    \eqno (10)
$$
     so that the matrix $\gamma$ has a null eigenvector. Differentiating
(10)    with respect to the squared edge lengths $g$ and contracting again
       with this null eigenvector gives
$$
0=\sum_{i\ne j} \sin{\theta\argsij} \pderiv{\theta\argsij}g \modsigma^3_i
\modsigma^3_j      \eqno (11)
$$
and since one can show by trigonometry that
$$
 \sin{\theta\argsij}  \modsigma^3_i \modsigma^3_j = {4\over3} \modsigma^4
\modsigma^2_{ij} \eqno (12) $$
  the result follows. \enddemo

\head      Independent Dihedral Angles \endhead

Unlike Palatini's action,  $J(g,\Theta)$ cannot be considered
as our first order action with $g$ and $\Theta$ the independent variables.
This is because, according to (7), the equation of motion one would get by
putting the variation with respect to $\Theta$ equal to zero would
imply that the area of every 2-simplex was zero, and so the metric of the
4-simplex would be very degenerate. However the
preceeding proof of equation (9) works because
for each 4-simplex there is a constraint amongst the $\theta$'s, or
in other words, the range
of the functions $\theta$ forms a submanifold of co-dimension one
in the space $\R^{10}$ of the $\Theta$'s for the 4-simplex.
Equation (9) shows that the
tangent vectors $w_i$ which lie in this surface
are the ones which satisfy $\sum_i w_i\modsigma^2_i =0$.

Therefore the ``first order'' stationary action
principle for Regge calculus can be stated as follows:

\definition{Restricted Variation} The action
 $$
J(g,\Theta)=\sum_{\sigma^2}|\sigma^2|(g)\left(2\pi-\sum_{\sigma^4>\sigma^2}
\Theta\args\right)                                       \eqno (13)
 $$
is varied subject to the condition that for each 4-simplex $\sigma^4$
the variables
$\left\{\Theta\argsij\right\}$ be the dihedral angles of a (different)
 4-simplex metric. \enddefinition

The ``equations of motion''
are the equations which give a stationary point of the action under this
variation. In (13), the area of a
2-simplex has been written $\modsigma^2 (g)$ to emphasise that this
is a function of the squared edge lengths $g$. The notation
$\metric(\sigma)$ denotes the space of metrics on a simplex; thus
$\metric(\sigma^4)\simeq\R^{10}$, and the positive definite ones form a
convex cone in this space.

Let us now compute the equations of motion from this action.
Introduce a second metric $\hat g\in \metric(\sigma^4)$ which is one
of the metrics for
which the variables $\left(\Theta\argsij\right)$ are, by
hypothesis, the dihedral angles. Let
$$\delta\Theta=\left(\delta\Theta{\args}_1,\delta\Theta\args_2,\ldots\right)$$
be a tangent vector in the space of
$\Theta$'s which preserves the constraints. Then, by (9), it
satisfies
 $$
0=\sum_{\sigma^2<\sigma^4}\modsigma^2 (\hat g)\delta\Theta\args \eqno(14)
$$
for each $\sigma^4$. Therefore, using equation (7), the action
$J(g,\Theta)$ is stationary for all such variations $\delta\Theta$
only if
$$
\modsigma^2 (g) = \mu\modsigma^2 (\hat g)  \eqno (15)
$$
for all 2-simplexes in the 4-simplex, with the constant of
proportionality $\mu$ fixed for each 4-simplex.
The other variational equation, for unrestrained variations of
the squared edge lengths $g$, is
$$0=\pderiv{J(g,\Theta)}g
= \sum_{\sigma^2}
\pderiv{|\sigma^2|}g
\left(2\pi-\sum_{\sigma^4>\sigma^2}\Theta\args\right)       \eqno (16)
$$
Clearly, one solution of (15) is that $g$ is proportional to $\hat
g$ on each $\sigma^4$, which
implies that $\Theta=\theta(g)$, and that (16) then reduces to the
usual Regge
equation of motion. In order to see if this is the only solution,
one has to examine (15), and determine whether or not the areas of
the 10 triangles of a 4-simplex fix the 10 edge lengths (and hence
the dihedral angles) of the simplex uniquely. The
situation is as follows.\medbreak

\proclaim{Proposition} There is an open convex region
$U\subset \metric(\sigma^4)\simeq \R^{10}$ of positive metrics for a
4-simplex, containing all the equilateral 4-simplexes, with the
property that if $(15)$ holds for $g,\hat g\in U$ and some $\mu>0$,
then $g$ is proportional to $\hat g$.
\endproclaim

\demo {Proof}
Define $\phi\colon \metric(\sigma^4)\rightarrow\R^{10}$ by
$\phi(g)=({\modsigma^2_{12} }^2 ,{\modsigma^2_{13} }^2,\ldots)$, the
squares of the 10 areas as a function of the squares of the 10 edge lengths.
Note that since $\phi$ is quadratic, condition (15) is equivalent to
$$\phi(g)=\phi(\mu^{1\over 2}\hat g)$$
and that if $g$, $\hat g$, satisfy this for some $\mu$ and are proportional,
then in fact $g=\mu^{1\over 2}\hat g$.

Therefore, it suffices to find an open convex set $U$ such that
$\phi$ is injective on $U$.
The Jacobian matrix $\d\phi$ can be computed at the metric which makes
each edge
length 1, i.e., an equilateral simplex. It is a multiple of
$$\pmatrix 0&0&0&0&0&0&0&1&1&1 \\
           0&0&0&0&0&1&1&0&0&1 \\
           0&0&0&0&1&0&1&0&1&0 \\
           0&0&0&0&1&1&0&1&0&0 \\
           0&0&1&1&0&0&0&0&0&1 \\
           0&1&0&1&0&0&0&0&1&0 \\
           0&1&1&0&0&0&0&1&0&0 \\
           1&0&0&1&0&0&1&0&0&0 \\
           1&0&1&0&0&1&0&0&0&0 \\
           1&1&0&0&1&0&0&0&0&0 \\ \endpmatrix ,
$$
which has determinant 48. The inverse function theorem assures us that there
is a convex neighbourhood of this equilateral simplex on which $\phi$
is injective. By taking all positive multiples of these metrics, one obtains
such a neighbourhood of all the equilateral simplexes.
\enddemo

Unfortunately, the set $U$ cannot cover all of the positive definite
metrics. This follows from an example which was discovered by Philip
Tuckey. Consider the one parameter family $(t,1,1,1,1,1,1,1,1,1)\in
\metric(\sigma^4)$. This is a family of metrics for which one edge has length
$\sqrt t$ and the rest have length 1.
The triangles in this 4-simplex all have edges lengths $(1,1,1)$ or $(1,1,
\sqrt t)$. The squared area of the latter is
$${1\over 8}t(4-t)$$
and so all areas are equal for parameters $t$ and $4-t$.

\head Extended Action  \endhead

Since the action principle outlined above is the constrained
variation of the function $J$, it is possible to use a Lagrange
multiplier technique to allow the variation to be completely
unconstrained. To do this, it is necessary to understand the nature
of the constraint among the dihedral angles. In turn, this will suggest
further ways in which the dihedral angles could be used as completely
independent variables.

The results are the following
\roster \item "(a)"
The constraint satisfied by the dihedral angles is that
the matrix
$$ \Gamma_{ij}=\cases -\cos\Theta_{ij} &i\ne j \\
1&i=j\endcases$$
 has exactly one eigenvector
$v_i$ of eigenvalue zero. Any set of numbers
\hbox{$\{0<\Theta_{ij}<\pi,i\ne j\}$}
satisfying this constraint is a set of dihedral angles for some non-degenerate
metric on a simplex, which may however not be positive definite.
\item "(b)" Two sets of edge lengths for a simplex have the same set of
dihedral angles only if one set is a scalar multiple of the other.
\endroster
In other words, metrics up to scale are equivalent to sets of angles
satisfying the constraint (a).

\demo {Proof of \rm(a)}
The previous section shows that dihedral angles of a
positive definite metric simplex do satisfy this constraint.
To prove the converse part of the statement,
let the (three dimensional) faces of a particular 4-simplex in
$\R^4$ be labelled $i=1,\ldots 5$. The $i$-th face determines a particular
trivector, an element of $\Lambda^3(\R^4)$, by the formula
$$\xi_i= {1\over 6}a_1 \wedge a_2\wedge a_3$$
with the three displacement vectors $a_1,a_2,a_3$ of the edges sharing any one
vertex of the face, arranged in the standard order dictated by an orientation
for
the simplex.
Since the faces form a closed boundary,
$$\sum_{i=1}^5\xi_i=0. \eqno (17)$$

Now let $\{\Theta_{ij}\}$ satisfy the conditions
and  $v_i$ be the null eigenvector of $\Gamma_{ij}$.
An inner product on $\Lambda^3(\R^4)$ is determined
by the formula
$$\left\langle\xi_i,\xi_j\right\rangle=v_iv_j\Gamma_{ij}.\eqno(18)$$
Since the matrix $\Gamma_{ij}$
has only one null eigenvector, this metric is non-degenerate, and determines
a metric on the simplex which induces this metric on $\Lambda^3(\R^4)$.
Indeed, in a standard basis, the components of one metric are just a
scalar multiple of the inverse of the matrix of the other. With this metric,
the volume of the $i$-th face is $|v_i|$.
\enddemo

\demo{Proof of \rm(b)} The only arbitrary choice in the previous proof
was the scaling of the eigenvector $v_i$ by a positive real number.
Therefore the metric on the
simplex is determined up to an arbitrary scaling.\enddemo

For the extended action, one needs to find a suitable function which is
zero on the constraint surface for the Lagrange multiplier. The obvious
choice is $\det(\Gamma_{ij})$.
Using this particular
function, the extended action principle is the following:

\definition {Extended Variation} The
equations of motion are given by the stationary points of the
unconstrained variation of the action
 $$
K(g,\Theta,\lambda)
= \sum_{\sigma^2}|\sigma^2|(g) \left(2\pi-\sum_{\sigma^4>\sigma^2}
\Theta\args\right) + \sum_{\sigma^4}\lambda(\sigma^4) \det\left(\Gamma_{ij}
(\Theta)\right)         \eqno   (19)
$$
with respect to $g,\Theta$, and
$\lambda=(\lambda(\sigma^4)_1,\lambda(\sigma^4)_2,\ldots)$, in the region
where
$g$ has positive signature.\enddefinition

\proclaim{Proposition}  The stationary points of $K$ have, in the region
where
$g$ is a positive definite metric, and $0<\Theta\args<\pi$, the same
values
of $g$ and $\Theta$ as the stationary points of the constrained
variation of
$J$ described above.\endproclaim

\demo{Proof} Variation with respect to $g$ clearly gives the same
equation, (16), as before. Variation with respect to $\lambda$ gives
the equations $\det(\Gamma_{ij})=0$ for each 4-simplex. As noted before,
the matrix $\gamma$, calculated from the dihedral angles $\theta$ of a
metric 4-simplex, has a null eigenvector. Hence $\det(\Gamma)=0$ is
certainly a necessary condition. A computation of the derivative of
$\det(\Gamma)$ with
respect to $\Theta_{ij}$ shows that it is zero if $\Gamma$ has two
independent null
eigenvectors, and then the variational equation for $\Theta$ becomes
$\modsigma^2 (g)=0$ for every 2-simplex in the 4-simplex. This
contradicts our hypothesis that $g$ is a positive definite metric. Thus we
are led to the conclusion that $\Gamma$ has only one independent null
eigenvector.
Let the null eigenvector be $V_i$, i.e. $\sum_i
\Gamma_{ij}V_i=0$. Computing the derivative of the determinant, one
finds that the variational equation for $\Theta_{ij}$ is
$$
0=\pderiv K{\Theta\argsij} = -\modsigma^2_{ij} (g) +
\kappa\lambda V_i V_j \sin \Theta_{ij}    \eqno (20)
 $$
where  $\kappa$ is a non-zero number, constant for each $\sigma^4$.
Since $\sin \Theta_{ij}>0$, and
the $\modsigma^2 $ are
all positive, it follows that either $V_i>0$ for all $i$, or
$V_i<0$ for
all $i$. Since the scaling of the eigenvector was arbitrary, we can
choose the sign so that they are all positive.

According to the results at the beginning of this section, the $\Theta_{ij}$
determine a metric, $\hat g$, for the 4-simplex, for which they are the
dihedral angles, and the $V_i$ the volumes of the faces. According to (12),
one can simplify (20) so that it becomes the previous equation of motion
(15), which asserts that the areas of the triangles computed with $g$
and $\hat g$ are proportional.
\enddemo

The use of the dihedral angles considered in this paper suggests some
further directions of study. With no restriction at all, a set of dihedral
angles specifies a constant curvature metric up to an overall scaling.
This suggests that it may be fruitful to consider a calculus based on
angles and constant curvature simplexes.

\enddocument